\documentclass[ showpacs,preprintnumbers,amsmath,amssymb]{revtex4}

\usepackage{amsmath,amsfonts}
 \usepackage{epsf}

\newcommand{\be}{\begin{equation}}
\newcommand{\ee}{\end{equation}}
\newcommand{\bes}{\begin{equation*}}
\newcommand{\ees}{\end{equation*}}
\newcommand{\bea}{\begin{eqnarray}}
\newcommand{\eea}{\end{eqnarray}}

\newcommand{\bra}{{\langle}}
\newcommand{\ket}{{\rangle}}

 \newcommand{\myfig}[3]{\begin{figure}[ht]
\begin{center}
\leavevmode \epsfxsize=#2cm \epsfbox{#1}
\end{center}
\caption{#3} \label{fig:#1}
\end{figure}}

\begin{document}

\title{Constraining Modified Gravity with Large  non-Gaussianities}
\author{Samuel E. V\'azquez}
\affiliation{$^1$ Perimeter Institute for Theoretical Physics, 31 Caroline St. North, Waterloo,
Ontario N2L 2Y5, Canada}
\begin{abstract}
In writing a covariant effective action for single field inflation, one is allowed to add a Gauss-Bonnet and axion-type curvature couplings. These couplings represent modifications of gravity, and are the unique higher-curvature terms that lead to second order equations of motion in four dimensions. 
  In this paper we study the observational consequences of such couplings for models with large non-gaussianities. Our focus is on the Gauss-Bonnet term.  In particular, we study an effective action where the scalar Lagrangian is a general function of the inflaton and its first derivative.   We show that, for large non-gaussianities,  one can write $f_{NL}$ in terms of only three parameters.  The shape of $f_{NL}$  is also studied, and we find that it is very similar to that of k-inflation.  We show that the Gauss-Bonnet  term 
enhances the production of gravitational waves, and allows a smaller speed of sound for scalar perturbations. This, in turn, can lead to larger non-gaussianities which can be constrained by observations. Using current WMAP limits on $f_{NL}$ and the tensor/scalar ratio, we put constraints on all parameters. As an example, we show that for DBI inflation, the Gauss-Bonnet coupling leads to an interesting observational window with both large $f_{NL}$ and a large amplitude of gravitational waves. Finally, we show that the Gauss-Bonnet coupling admits a de-Sitter phase with a relativistic dispersion relation for scalar perturbations.

\end{abstract}

\maketitle

\section{Introduction}

Cosmology has entered an era of unprecedented progress. High precision measurements of the cosmological parameters have led to a coherent picture of the history of our universe that seems to favor the inflationary paradigm \cite{Guth}. Moreover, a future detection of large
non-gaussianity in the cosmic microwave background (CMB) would falsify the simplest inflationary scenario, namely, single field slow-roll inflation \cite{Komatsu, Komatsu1, nongauss}.  

On the theoretical side, there has been great activity in trying  to produce large non-gaussianities in single and multiple-field inflationary models.  For  single field inflation, large non-gaussianities are easiest to produce in models with a small speed of sound (see e.g. \cite{Kach, largefnl}).  On a parallel set of developments, there has been recent interest in developing a systematic effective field theory of single field inflation \cite{Senatore, Weinberg}. In ref. \cite{Senatore}, such approach was applied directly to the Lagrangian describing the perturbations around the inflationary solution. The effective action can be viewed as an expansion  in powers of $(g^{00} +1)$ and the extrinsic curvature $K_{a b}$ of the constant time hypersurfaces. Such approach is quite general, and provides a straightforward way of calculating all CMB observables directly from the effective action for the fluctuations.

On the other hand, one would like to understand how the various terms in the effective action for the fluctuations relate to the effective action of the inflaton itself. A method to build such an effective action was introduced by Weinberg in \cite{Weinberg}.  In this approach, one considers all marginal and irrelevant operators involving the inflaton and the metric. Among these terms, there are higher curvature invariants coupled to the inflaton. Generically, such terms will contain higher time derivatives on the fields which need to be eliminated using the first order equations of motion. Otherwise, one would be propagating more degrees of freedoms than intended. Weinberg showed that, after such eliminations and to leading order in the derivative expansion, the resulting action for the inflaton takes the familiar k-inflation type form plus two extra couplings between the inflaton and the Weyl tensor.  Such extra couplings can be written instead in terms of the Gauss-Bonnet tensor and an axion-type coupling. In this way, one has an effective action that leads to second order equations of motion explicitly.

One can then ask if it is possible to re-sum such an expansion. Moreover, one would like to write down a general {\it local} action for the inflaton coupled to gravity that leads to second order equations of motion for all fields. By abuse of notation we will call such re-summation a ``UV completion" of the effective theory. An advantage of having such action is that one can then build directly the low energy effective action for the fluctuations as in \cite{Senatore}, have a clear physical interpretation of the various couplings, and a better assessment of their relative importance. 
If  we also insist in preserving general covariance in the UV completion, it is not hard to see that an obvious candidate for such action is \footnote{We cannot claim that this is the most general action that propagates only the spin two graviton and a scalar degree of freedom.  Nevertheless, most alternative  constructions will either break general covariance or involve another scalar field, even if it is not dynamical (see e.g. \cite{cuscuton}).}
\be 
\label{S}
S =  \int \sqrt{-g}\left[ \frac{1}{2}R + P(X,\phi) + V_1(\phi) E_4 +  V_2(\phi) \epsilon^{abcd} R_{ab}^{\;\;\; ef} R_{cdef}\right]\;.\ee
where $E_4$ is the Gauss-Bonnet combination,
\be E_4 = R_{abcd}R^{abcd} - 4 R_{ab} R^{ab} + R^2\;,\ee 
and $X = -\frac{1}{2} \nabla_a \phi \nabla^a \phi$ is the kinetic term. In writing (\ref{S}) we assume that we work in the Einstein frame. The second term in the action is the familiar k-inflation type \cite{kinf}.  The last two couplings in the effective action would be topological invariants in four dimensions if both potentials $V_1,V_2$ were constant. This is the reason why they lead to second order equations of motion for general $V_i$. Note that if these potentials depended on $X$,  one would end up with equations of motion depending on more than two time derivatives.

The last two terms in the action (\ref{S}) represent modifications of Einstein's gravity, and as such, they have been studied numerous times (see e.g. \cite{modified}). Moreover, such couplings are known to arise in string theory \cite{threshold}. 
The last term in (\ref{S})
is of the axion-type and it was studied long ago in \cite{hwang1}, where it was shown that such coupling does not affect the evolution of scalar fluctuations to quadratic order. We have verified that this is still true to cubic order. Therefore, we discard this coupling in what follows \footnote{Note, however, that the axion-type coupling does affect the gravitational waves by giving an extra helicity dependence to the tensor power spectrum \cite{Weinberg, hwang1}.}.
The Gauss-Bonnet term, on the other hand, has been studied many times in the context of Dark Energy (e.g. \cite{modified}).  For other studies in the context of early cosmology see \cite{Sato}.   This term does contribute to scalar fluctuations and it is the main focus of this work.

The purpose of this paper is to study the observational signatures of the Gauss-Bonnet coupling in the context of inflation. Moreover, our main interest will be in models will large non-gaussianities.  In the following, we compute the non-gaussianity parameter $f_{NL}$ using the action (\ref{S}) in the limit of a small speed of sound. Moreover, we perform such calculation to leading order in the slow roll parameters, but to all orders in the ``strength" of the Gauss-Bonnet coupling, defined as
 \be \label{epsilon1}g \equiv 8 V_1'(\phi) H \dot\phi\;,\ee
 where $H$ is the Hubble parameter.
 In making the calculations, we assume that the parameter $g$ is slowly varying in time.
  We find that for large-nongaussianities,  $f_{NL}$ can be written in terms of only three parameters. We study the shape of of $f_{NL}$ as a function of these parameters.   We find that the shape of $f_{NL}$ is always very close to that of k-inflation, even in the limit $g \rightarrow \infty$.  We discuss in which cases deformations from this shape might be observed. 
  
 The spectrum of gravitational waves is also studied, and we find an enhancement due to the Gauss-Bonnet term.  Using WMAP limits on the equilateral  $f_{NL}$ and the tensor/scalar ratio, we  put constraints on the different parameters.   For the particular case of DBI inflation, we show that one is left with only a two-parameter family of $f_{NL}$. In this case, one can put a more precise constraint on the Gauss-Bonnet coupling:
 \bes g_\text{DBI} \lesssim 3\;.\ees
 We also discuss implications of the Gauss-Bonnet coupling on the Lyth bound in the context of DBI inflation \cite{Lyth}. 
   
  An interesting aspect of the Gauss-Bonnet coupling is that scalar perturbations are non-trivial even in a de-Sitter background, just like the do in the Ghost Condensate \cite{ghost}.  However, we find that in our case, quadratic scalar fluctuations have the familiar relativistic dispersion relation $\omega \sim k$ instead of the non-relativistic one $\omega \sim k^2$ of the Ghost Condensate. This matches perfectly with the new de-Sitter limit found in \cite{Senatore} using the effective action of the fluctuations. Therefore, the Gauss-Bonnet coupling is precisely the modification of gravity that leads to such limit.

  The paper is organized as follows. In section II we study the equations of motion for the background solution and the quadratic scalar fluctuations. We also point out the different limits used in the calculations. In section III we calculate $f_{NL}$ and study its shape.  In section IV we calculate the gravitational wave spectrum. We discuss the various constraints on the parameter space.  In section V we study the case of DBI inflation. Finally, we close with some final comments and future directions in section VI.

 \section{Background Solution and Quadratic Fluctuations}

The equations of motion for the homogeneous background that follow from the action (\ref{S}) can be written as,
\bea \label{eom1}E &=& 3 H^2 (1 + g)\;, \\
 \label{eom2}\dot E &=&- 3 H(E+ P) + 3 H^3(1 - \epsilon) g\;,
 \eea
 where
 \bes E = 2 X \partial_X P - P\;,\;\;\;\; X = \frac{1}{2} \dot\phi^2\;,\;\;\;\;\epsilon = -\frac{\dot H}{H^2}\;,\ees
and $g$ is defined in Eq. (\ref{epsilon1}).

Since we are interested in isolating the contribution to $f_{NL}$ of the Gauss-Bonnet coupling, we will set $\epsilon= 0$. That is, we will work in a de-Sitter background \footnote{One does not really need to assume $\epsilon = 0$ but only that $g$ is much larger than $\epsilon$ or any of the slow roll parameters defined below.}. The existence of such limit will be established in the next section. Moreover, we will assume that the parameter $g$ varies slowly with time. Therefore, to the first approximation we can consider it to be a constant. Note however, that $g$ itself can be large. Therefore, we will do our calculations to all orders in $g$. 

Under these assumptions, one can easily show from (\ref{eom1}) and (\ref{eom2}) that,
\bes X\partial_X P = \frac{1}{2} H^2 g\;.\ees
This will be a very useful relation in what follows. The speed of sound of this model is given by,
\be 
c_s^2 = \left(1 + 2 X \frac{\partial^2_X P}{\partial_X P}\right)^{-1}\;.\ee
Since we are interested in the limit of large non-gaussianities, we will be working with a small speed of sound. In this case, one can eliminate time derivatives of $\phi$ by
\bes \frac{\dot \phi^2}{H^2} \approx \frac{\alpha^2}{c_s^2}\;,\ees
where we have defined
\be \label{alpha} \alpha^2 \equiv \frac{\partial_X P}{\partial_X^2 P H^2}\;.\ee
In the limit of small speed of sound, we do not need to assume that $\dot\phi/H \gg1$, but only that $\dot \phi/H$ is much larger that any of the slow roll parameters. By slow roll parameters we mean any of the parameters that encode the time evolution of the solution such as $\epsilon$, $\eta \equiv \dot\epsilon/ H\epsilon$, etc.

In this case, one can show from the equations of motion (\ref{eom1}) and (\ref{eom2})  that derivatives with respect to the field $\phi$ are suppressed, e.g.
\bes \frac{\partial_\phi P}{H^2} \sim \frac{(\text{slow roll})}{\dot \phi /H} \ll 1\;.\ees
Similar limits can be shown for other quantities involving derivatives of the scalar field like, e.g., $\partial_\phi \partial_X P$.

Let us now define the parameters
\be \label{slowroll1}\eta_1 \equiv V_1''(\phi) \dot\phi^2\;,\;\;\;\; \xi = \frac{\dot X}{2 H X} \equiv \frac{\ddot \phi}{ H \dot\phi}\;.\ee
 Note that $\xi$ is a slow roll parameter, and by assumption, it must be small. It is then easy to show that,
\be \label{slowroll2} \frac{\eta_1}{g} = \frac{\dot g}{H g} + \epsilon - \xi \ll1\;.\ee
Therefore, in the slow roll limit, $\eta_1$ is small compared to $g$. This means that we will be able to ignore higher derivatives of the potential $V_1'(\phi)$ in the following.

Note that the slow roll conditions are necessary to ensure an almost scale-invariant spectrum of scalar fluctuations. Nevertheless, we will see that the value of $g$ does not affect the scalar tilt, and hence it can be much larger than any of the slow roll parameters. 

\subsection{Quadratic Fluctuations}
To derive the action for the fluctuations we will use the ADM formalism, where we write the metric as
\bes ds^2 = -N^2 dt^2 + h_{i j} (dx^i + N^i dt)(dx^j + N^j dt)\;.\ees
It turns out to be technically simplest to work in the gauge:
\be\label{gauge} \delta \phi = \varphi\;,\;\;\; h_{ij} = e^{2 \sigma}(\delta_{ij} + \gamma_{ij})\;,\;\;\; \partial_i \gamma_{ij} = 0\;,\;\;\; \gamma_{ii} = 0\;.\ee This gauge is different from the unitary gauge $\delta \phi = 0$ used in \cite{Senatore}. However, we will comment on their relation in due course.

To relate this gauge choice to the physical (conserved) curvature perturbation, we can use the $\delta {\cal N}$ formalism \cite{deltaN}. In this formalism, one can relate the gauge invariant curvature perturbation $\zeta$ to the number of e-foldings since the time of horizon crossing ($t_*$)
\bes \zeta(t,\vec x) = {\cal N}(t,\vec{x}) - {\cal N}_0(t)\equiv \delta {\cal N}\;,\ees
where 
\be\label{Nn} {\cal N}(t,\vec x) = \int_{t_*}^t H(t',\vec{x}) dt'\;,\ee
and ${\cal N}_0$ denotes the background value without the perturbation.  We can then view ${\cal N}$ as a function of the scalar field perturbation evaluated at the time $t_*$.  We  can then write,
\be \label{Nexp} \delta {\cal N} = \frac{\partial {\cal N}}{\partial \phi_*} \varphi_* + \frac{1}{2} \frac{\partial^2 {\cal N}}{\partial \phi_*^2} \varphi_*^2 + \ldots \ee 
The first derivative follows directly from (\ref{Nn}):
\bes \frac{\partial {\cal N}}{\partial \phi_*}   = - \frac{ H_*}{\dot \phi_*} \;,\ees 
where, as usual, the star denotes evaluation at $t_*$. 

Higher derivatives of ${\cal N}$ will be suppressed by powers of the slow roll parameters. For example, by using the equations of motion (\ref{eom1}) and (\ref{eom2}), and  in the limit where $\epsilon = 0$, one can show that
\bes \partial_{\phi} \left(\frac{H}{\dot\phi}\right) \approx \frac{4}{3} \left(\frac{H}{\dot\phi}\right)^2 \frac{g\xi}{c_s^2} + {\cal O}\left(\frac{1}{c_s}\right)\;.\ees
  It is then easy to show that the extra terms in the expansion (\ref{Nexp}) will give corrections to $f_{NL}$ in powers of $\sim (\text{slow roll})/c_s^2$. We assume that the slow roll parameters are small such that $(\text{slow roll})/c_s^2 \ll 1$. In any case, we will be working in a de-Sitter background where these corrections are exactly zero.

The two point function of the curvature perturbation is given by,
\be \label{scalarbi} \bra \zeta_{\vec k}(t) \zeta_{\vec k'} (t)\ket \approx \frac{H_*^2}{\dot \phi_*^2} \bra \varphi_{\vec k}(t_*) \varphi_{\vec k'} (t_*)\ket \equiv (2 \pi)^3 \delta^{(3)}(\vec k + \vec k') P_\zeta(\vec{k})\;.
\ee
Similarly, the three point function can be written as
\bea  \label{trispectrum} \bra \zeta_{\vec k_1}(t)\zeta_{\vec k_2}(t)\zeta_{\vec k_2}(t)\ket &\approx& - \frac{H_*^3}{\dot \phi_*^3}
\bra \varphi_{\vec k_1}(t_*) \varphi_{\vec k_2}(t_*) \varphi_{\vec k_3}(t_*)\ket    \nonumber \\
&\equiv& (2 \pi)^3 \delta^{(3)}(\vec k_1+\vec k_2+\vec k_2)\left(- \frac{6}{5} f_{NL} \right)\sum_{i < j} P_\zeta(\vec k_i) P_\zeta(\vec k_j)\;. \eea
In the last line we have defined the non-gaussianity parameter $f_{NL}$, following the conventions in \cite{shape}.

To linearized order, the solution of the constraint equations can be parametrized by two scalars,
\bes N = 1 + \delta N\;,\;\;\; N^i = \partial_i \chi\;.\ees
After many integrations by parts, the quadratic Lagrangian for the Einstein-Hilbert, scalar and Gauss-Bonnet part of the action read respectively,
\bea 
\label{quad1}
S_\text{EH}^{(2)} &=& \frac{1}{2} \int e^{3\sigma} H \delta N\left( 3 H \delta N + 2\nabla^2 \chi\right)\;,\\
S_{\phi}^{(2)} &=& \frac{1}{2} \int e^{3\sigma} \left[ \partial_\phi^2 P \varphi^2 + 2 X_2 \partial_X P + 2 \delta N\left(\partial_\phi P \varphi + X_1 \partial_X P\right) + X_1\left(2 \partial_\phi \partial_X P \varphi + X_1 \partial^2_X P\right)\right]\;,\\
\label{quad3}
S_\text{GB}^{(2)} &=& - \int e^\sigma 8 V_1' H^2 \delta N \nabla^2 \varphi  + \int e^{3\sigma} \left\{ -48 V_1' H^3 \dot \phi \delta N^2 + 12(H^2 + \dot H) H^2 V_1'' \varphi^2 \right.\nonumber \\
 &&\left. -8 H^2 \nabla^2 \chi \left[ \varphi (V_1' H  - \dot\phi V_1'') - V_1' \dot \varphi\right]  + 24 H^2 \delta N \left[ \varphi V_1'' H \dot\phi + V_1' (-\dot\phi \nabla^2 \chi + H \dot\varphi)\right]\right\}\;.
\eea
where
\bea X_1 &=& \dot \phi (\dot \varphi - \dot \phi \delta N)\;, \\ 
\label{X2}
X_2 &=& \frac{1}{2}\left[ 3 \dot\phi^2 \delta N^2  - e^{-2\sigma} (\nabla \varphi)^2 - 2 \dot \phi \partial_i \varphi \partial_i \chi - 4 \dot\phi \delta N \dot\varphi + \dot\varphi^2\right]\;.\eea
No approximations have been made in deriving Eqs. (\ref{quad1}) - (\ref{X2}).

As pointed out in the previous section, in the slow roll limit we can ignore the terms involving the second derivative of the potential $V_1$. 
Varying Eqs. (\ref{quad1}) - (\ref{quad3}) with respect to the constraints we obtain,
\bea \label{N} \delta N &\approx& \frac{c_s g}{2 \alpha(1 + \frac{3}{2}g)} \frac{\dot\varphi}{H}\;,\\
 \label{chi}
 \nabla^2 \chi &\approx& -\left(1 + \frac{3}{2}g\right)^{-1}\left[ \frac{g c_s}{2\alpha H} e^{-2\sigma} \nabla^2 \varphi + \frac{g}{2\alpha c_s} (\dot\varphi - \dot\phi \delta N)\right]\;,\eea
 where we have kept only the leading terms in the limit $c_s \rightarrow 0$ \footnote{Note however, that terms involving $e^{-2\sigma} c_s$ will give ${\cal O}(1/c_s^2)$ contributions to $f_{NL}$.}.
 Therefore, we see that in this limit,
 \be \label{est} \delta N \sim c_s\;,\;\;\; \chi \sim \frac{1}{c_s}\;.\ee
 This order-of-magnitude estimate is important to determine which terms in the action survive the small speed of sound limit.
 
 Inserting the solution for the Lagrange multipliers (\ref{N}) and (\ref{chi}) back into the quadratic actions (\ref{quad1}) - (\ref{quad3}) and taking the small speed of sound limit, we obtain the quadratic action for the scalar fluctuations:
\be \label{S2}\lim_{c_s \rightarrow 0} S^{(2)} = \int e^{3\sigma} f(g)\left[ \dot\varphi^2 - \tilde c_s^2 (\nabla\varphi)^2 e^{-2\sigma}\right]\ee
where
\bea f(g) &=& \frac{2 g (1 + g)^2}{\alpha^2(2 + 3g)^2}\;, \nonumber \\
\tilde c_s^2 &=& \frac{(1 + 2 g)(2 + 3 g)} {2(1 + g)^2}c_s^2\;. \nonumber \eea
Note that the quadratic action for the fluctuations is non-trivial even in the de-Sitter background that we are considering. Moreover, we see that $g$ must be positive in this limit in order to give the correct sign for the kinetic term in Eq. (\ref{S2}).

The dispersion relation for the quadratic fluctuations is precisely of the relativistic type $\omega \sim k$. Such fluctuation spectrum around de-Sitter was first found in \cite{Senatore} by studying the effective action for the scalar fluctuations directly. We then see that our approach gives a physical interpretation to such fluctuations: they are generated by the Gauss-Bonnet coupling of the inflaton.
The precise dictionary between our variables and those of Ref. \cite{Senatore} is the following: $\zeta = - H \pi = (H/\dot \phi) \varphi$, where $\pi$ is the ``Goldstone Boson" of \cite{Senatore}.
Moreover, in our gauge, the extrinsic curvature and the Lapse can be written as
\bes \delta K_{i j} = -  \partial_i \partial_j \chi\;,\;\;\; \delta N \approx -\frac{1}{2} (1 + g^{00})\ees
It is then easy to see that the terms that give the relativistic dispersion for the quadratic fluctuations come precisely from couplings of the form $(1 + g^{00}) \delta K^i_i$ in the effective action. These were the terms studied in \cite{Senatore}.

The quantization of the perturbations using the action (\ref{S2}) proceeds in the standard way. We write,
\bes \varphi(\tau,\vec{x}) = \int\frac{d^3 k}{(2\pi)^3} \varphi_{\vec{k}}(\tau) e^{i \vec{k}\cdot \vec{x}}\;,\ees
where the Fourier transform is written in terms of the standard harmonic oscillator operators
\bes \varphi_{\vec{k}}(\tau) = u_{\vec{k}}(\tau) a_{\vec k}^\dagger + u_{\vec{k}}^*(\tau) a_{-\vec k}\;.\ees 
Here we are using conformal time defined by $dt = e^{\sigma} d\tau$. For de-Sitter space, $e^\sigma = - 1/(H \tau)$ where $\tau \in (-\infty,0]$.

The properly normalized Bunch-Davies vacuum is given by \cite{Davies},
\bes u_{\vec{k}} = \frac{H}{\sqrt{4 f\tilde c_s^3 k^3}} (1- i k \tilde c_s \tau) e^{i k \tilde c_s \tau}\;.\ees
The power spectrum follows from the definition (\ref{scalarbi}):
\be\boxed{\label{Pscalar} P_{\zeta}(k) = \frac{H^2 (2 + 3 g)^2}{ 8 g (1 + g)^2 \tilde c_s }  \frac{1}{k^3}\equiv { \cal P_\zeta} \frac{1}{k^3}\;.}\ee
Note that, in the slow roll limit,   we have a scale invariant spectrum. Therefore, observations of the scalar tilt do not put constraints on the value of the Gauss-Bonnet coupling.

\section{Non-Gaussianities}
To study the non-gaussianities for this model, we need to expand the action to cubic order.  In this section we will show only the leading terms in the limit of a small speed of sound. Moreover, we will always ignore the mixing with gravity. To estimate the size of the various terms, the limits (\ref{est}) are useful. It turns out that the cubic action is of order $\sim 1/c_s$.  There are many integrations by parts  necessary to put the results in a simple form. The final result for the cubic couplings takes the form,
\bea \label{cubic1}\lim_{c_s \rightarrow 0} S^{(3)}_\text{EH} &=& -\frac{1}{2} \int e^{3 \sigma} \delta N\left[ \partial_i \partial_j \chi \partial_i \partial_j \chi - (\nabla^2\chi)^2\right]\;, \\
\label{cubic2}\lim_{c_s \rightarrow 0} S^{(3)}_\phi &=& \int e^{3 \sigma} \left( X_1 X_2 \partial_X^2 P + \frac{1}{3} X_1^3 \partial_X^3 P\right)\;,\\ 
\label{cubic3}\lim_{c_s \rightarrow 0} S^{(3)}_\text{GB} &=& - \int e^\sigma  \frac{g c_s }{\alpha H} \delta N \left[ \partial_i \partial_j \varphi \partial_i \partial_j \chi - (\nabla^2 \varphi)(\nabla^2 \chi)\right] - \frac{3}{2} \int e^{3 \sigma} g \delta N \left[ \partial_i \partial_j \chi \partial_i \partial_j \chi - (\nabla^2\chi)^2\right] \nonumber \\
&& - \int e^{3\sigma} \frac{g c_s}{\alpha} \partial_i \varphi \partial_i \chi \nabla^2 \chi + \frac{1}{2} \int e^{ 3\sigma} \frac{g c_s}{\alpha H} \dot\varphi \left[ \partial_i \partial_j \chi \partial_i \partial_j \chi - (\nabla^2\chi)^2\right] \;,\eea
where,
\bea
X_1 &=&  \frac{\alpha H}{c_s} \left( \dot \varphi - \frac{\alpha H}{c_s} \delta N\right)\;, \nonumber \\
 \lim_{c_s \rightarrow 0} X_2 &=&  - \frac{1}{2} e^{-2 \sigma} (\nabla \varphi)^2 -  \frac{\alpha H}{c_s} \partial_i \varphi \partial_i \chi \;. \nonumber 
 \eea
 
 In the usual k-inflationary scenario ($g = 0$), the Einstein-Hilbert action Eq. (\ref{cubic1}) gives a contribution to the tri-spectrum which is of higher order in slow roll:  $f_{NL} \sim \epsilon^2/c_s^2$. Therefore, this term has been ignored in previous calculations. However, with a non-zero Gauss-Bonnet coupling, we have seen that fluctuations can exist even in a de-Sitter background. Therefore one can have $g \gg \epsilon$. In this case, which is the main focus of this paper, all terms in Eqs. (\ref{cubic1}) - (\ref{cubic3}) are equally important. 

In the language of the effective field theory of \cite{Senatore}, we can readily identify some of the cubic couplings. For example, the Einstein-Hilbert term can be written as $S^{(3)}_\text{EH} \sim \int (1 + g^{00}) (\delta K^i_j \delta K_i^j - (\delta K_i^i)^2)$. However, since our gauge choice is different from \cite{Senatore}, not all terms in (\ref{cubic1}) - (\ref{cubic3}) can be written only in terms of the Lapse and the extrinsic curvature. 

In order to calculate the scalar three point function we follow the standard procedure. To leading order in the cubic perturbation, the three point function is given by
\bea \bra \varphi_{\vec k_1}(t) \varphi_{\vec k_2}(t) \varphi_{\vec k_3}(t)\ket &=& - i \int_{-\infty}^t dt' \bra 0|[\varphi_{\vec k_1}(t) \varphi_{\vec k_2}(t) \varphi_{\vec k_3}(t), H_\text{int}(t')]|0\ket \nonumber \\
&\approx& - i \int_{-\infty}^0 d\tau' e^{\sigma(\tau')}\bra 0|[\varphi_{\vec k_1}(0) \varphi_{\vec k_2}(0) \varphi_{\vec k_3}(0), H_\text{int}(\tau')]|0\ket\;, \nonumber \eea
where in the last line we have changed the integration to conformal time, and made the usual late time approximation. The interaction Hamiltonian is given in terms of the cubic Lagranian density: $H_\text{int}  =- {\cal L}^{(3)}$.

The cubic Lagrangian density can be written in Fouriers space as,
\bes {\cal L}^{(3)}(t) = \int \frac{d^3 k_1}{(2\pi)^3}\frac{d^3 k_2}{(2\pi)^3}\frac{d^3 k_3}{(2\pi)^3} (2\pi)^3 \delta^{(3)}(\vec{k}_1 +\vec{k}_2+\vec{k}_3) \tilde {\cal L}^{(3)}(k_1,k_2,k_3; t)\ees
Then, it is not hard to see that the three point function can be written as,
\bea \bra \varphi_{\vec k_1}(t) \varphi_{\vec k_2}(t) \varphi_{\vec k_3}(t)\ket  =  - i (2 \pi)^3 \delta^{(3)}(\vec{k}_1 +\vec{k}_2+\vec{k}_3) \prod_{i = 1}^3 |u_{\vec{k}_i}(0)|^2 \int_{-\infty}^0 \frac{dx}{ H x} \tilde {\cal L}^{(3)}(k_1,k_2,k_3; x) + \text{c.c.} + \text{perms.} \nonumber \eea
where we have defined the integration variable $x =\tilde c_s \tau$. Moreover, $\tilde {\cal L}^{(3)}(k_1,k_2,k_3; x)$ is calculated by taking the cubic Lagrangian density that follows from Eqs. (\ref{cubic1}) - (\ref{cubic3}) and replacing the fields $\varphi$ using the rules:
\bes \varphi \rightarrow (1- i k x)e^{i k x} \;,\;\;\; \dot\varphi \rightarrow - H k^2 x^2 e^{i k x}\;,\ees
where $k= |\vec{k}|$.
Moreover, all spatial derivatives are replaced by their fourier transform: $\partial_i \rightarrow i (\vec{k})_i$. Using the definition of $f_{NL}$ in Eq. (\ref{trispectrum}), we can write
\be \label{fnllag} f_{NL}(k_1,k_2,k_3) =- \frac{5 \alpha^3 {\cal P}_\zeta }{6 c_s^3 \sum_i k_i^3}\left[i\int_{-\infty}^0 \frac{dx}{ H x} \tilde {\cal L}^{(3)}(k_1,k_2,k_3; x) + \text{c.c.} + \text{perms.}\right]\;.\ee

In doing the integrals, one projects to the correct interacting ground state by rotating slightly the contour so that $\tau \rightarrow (1 - i \epsilon) \tau$. With these considerations, it is straightforward to calculate the three point function. As an example, let us work out the Einstein Hilbert term, Eq. (\ref{cubic1}). The Fourier transform Lagrangian density can be written as,
\bes \tilde {\cal L}^{(3)}_\text{EH}(k_1,k_2,k_3; x) = - \frac{1}{2}  \left(- \frac{\tilde c_s }{H x}\right)^3\left[(\vec{k_2} \cdot \vec{k_3} )^2- k_2^2 k_3^2\right] \tilde{\delta N}(k_1) \tilde \chi(k_2) \tilde \chi(k_3) e^{i K x}\;,\ees
where,
\bes  \tilde{\delta N}(k) =  - \frac{ \tilde c_s a b}{2 \alpha} k^2 x^2 \;,\;\;\;\; \tilde \chi(k) = \frac{b H x^2}{4 a \alpha \tilde c_s}\left[-2 + b -2 a^2(1  - i k x)\right]\;,\ees
and
\bes c_s = a \tilde c_s\;,\;\;\; K = k_1 + k_2 + k_3\;,\;\;\;\; a^2 = \frac{2(1 + g)^2}{(1 + 2 g)(2 + 3 g)}\;,\;\;\; b = \frac{g}{1 + \frac{3}{2} g}\;.\ees
Therefore, the time integration gives, 
\bea \int_{-\infty}^0 \frac{dx}{ H x} \tilde {\cal L}_\text{EH}^{(3)}(k_1,k_2,k_3; x) &=&-\frac{i b^3 \tilde c_s^2 k_1^2 }{32 a H^2 K^5 \alpha^3}\left[(\vec{k_2} \cdot \vec{k_3} )^2- k_2^2 k_3^2\right] \nonumber \\
&&\times \left(48 k_2
   k_3 a^4+6 \left(2 a^2-b+2\right) K (k_2+k_3) a^2+\left(-2 a^2+b-2\right)^2
   K^2\right)\;. \nonumber 
   \eea
   Using the definition (\ref{fnllag}) we find the contribution to $f_{NL}$ of the Einstein-Hilbert action:
   \bea \label{fnlEH}
   f_{NL}^{(\text{EH})} &=& \frac{5 b^2 (3 b-2) k_1^2 \left[(\vec{k}_2 \cdot \vec{k}_3)^2-k_2^2 k_3^2\right] }{96 a^4 (b-2)^2 \tilde c_s^2 K^5 \sum_i k_i^3}\left\{6 \left[\left(2 a^2-b+2\right) K
   k_3 \right. \right. \nonumber \\
   &&\left.\left.+k_2 \left(8 k_3 a^2+\left(2 a^2-b+2\right) K\right)\right]a^2+\left(-2 a^2+b-2\right)^2
   K^2\right\} + \text{perms.}\eea
   
   In calculating the contribution from the scalar sector, one encounters the question of how big is $\partial_X^3 P$ compared to $\partial_X^2 P$ (note that both $X_1 X_2$ and $X_1^3$ scale as $\sim 1/c_s^3$). There is no way of knowing this without a detailed form of $P$. We will simply introduce a new variable $\lambda$ and write,
  \be \label{lambda}\lambda \equiv \alpha^2 H^2 \frac{\partial_X^3 P}{\partial_X^2 P} = \frac{\partial_X P \partial_X^3 P}{(\partial_X^2 P)^2}\;.\ee
  Then, following the same steps as in the Einstein-Hilbert term, it is easy to show that the contribution from the scalar sector to $f_{NL}$ is given by
   \bea \label{fnlphi}f_{NL}^{(\phi)} &=& \frac{5 (2 - b) \lambda  k_1^2 k_2^2 k_3^2}{6 a^4 \tilde c_s^2 K^3 \sum_i k_i^3} \nonumber \\
   &&+
   \frac{5 (\vec{k}_2 \cdot \vec{k}_3) k_1^2 }{12 a^4 (2 - b)\tilde c_s^2 K^3 \sum_i k_i^3}\left[2 (b-1) \left(K+2 k_2\right) k_3 a^2+\left(2 a^2 (b-1)-(b-2) b\right) K
   \left(K+k_2\right)\right]\nonumber \\
   &&+ \text{perms.}
   \eea
   
   Finally, the contribution from the Gauss-Bonnet term gives,
    \bea \label{fnlGB} f_{NL}^{(\text{GB})} &=& -\frac{5 b^2 }{48 a^4 (b-2)^2 \tilde c_s^2 K^5 \sum_i k_i^3} \left\{(b-2) \left((\vec{k}_2 \cdot \vec{k}_3)^2-k_2^2 k_3^2\right) \left[18 k_2 \left(8 k_3 a^2+\left(2 a^2-b+2\right) K\right) a^2 \right. \right. \nonumber \\
    &&\left. \left.+\left(6 a^2-3
   b+2\right) K \left(6 k_3 a^2+\left(2 a^2-b+2\right) K\right)\right] k_1^2 \right. \nonumber \\
   && \left. +2 K (\vec{k}_1 \cdot \vec{k}_2) k_3^2  k_1 \left[4 \left(\left(2 a^2-b+2\right) K k_3+k_2
   \left(6 k_3 a^2+\left(2 a^2-b+2\right) K\right)\right) a^2+\left(-2 a^2+b-2\right)^2 K^2\right] \right.\nonumber \\
   &&\left.+2 K^2 (\vec{k}_1 \cdot \vec{k}_2) k_3^2 \left[2
   \left(\left(2 a^2-b+2\right) K k_3+k_2 \left(4 k_3 a^2+\left(2 a^2-b+2\right) K\right)\right) a^2+\left(-2 a^2+b-2\right)^2
   K^2\right]\right\}\nonumber \\
   &&+ \text{perms.} \eea   
   The total contribution to $f_{NL}$ is then
   \be\label{fnlfinal} \boxed{ f_{NL} = f_{NL}^{(\text{EH})} + f_{NL}^{(\phi)} + f_{NL}^{(\text{GB})}\;,}\ee
   where the different terms are given in Eqs. (\ref{fnlEH}), (\ref{fnlphi}) and (\ref{fnlGB}).
   Note that the wave-vector dependence of $f_{NL}$ can be written in terms of the magnitudes $k_i$ by using,
   \bes \vec{k}_i \cdot \vec{k}_j = \frac{1}{2} (k_l - k_i - k_j)\;,\;\;\; l \neq i,j\;.\ees
   
   If we take the limit of vanishing Gauss-Bonnet coupling $g \rightarrow 0$, only the scalar contribution survive and we get,
   \bea \label{fnllim} \lim_{g \rightarrow 0} f_{NL}  &=& \frac{5 \lambda  k_1^2 k_2^2 k_3^2}{3 c_s^2 K^3 \sum_i k_i^3} + \frac{5 k_1^2 \left(-k_1^2+k_2^2+k_3^2\right) }{24 c_s^2 K^3
   \sum_i k_i^3}\left[K \left(K+k_3\right)+k_2 \left(K+2 k_3\right)\right] + \text{perms.} \nonumber\\
   &=& \frac{5 \lambda  k_1^2 k_2^2 k_3^2}{3 c_s^2 K^3 \sum_i k_i^3} -\frac{10}{3 c_s^2 \sum_i k_i^3}\left[ \frac{3k_1^2 k_2^2 k_3^2}{2 c_s^2 K^3 }  - \frac{1}{K} \sum_{i < j} k_i^2 k_j^2 + \frac{1}{2 K^2} \sum_{i \neq j} k_i^2 k_i^3 + \frac{1}{8} \sum_i k_i^3 \right]\eea
  One can check that Eq. (\ref{fnllim}) coincides with the $f_{NL}$ calculated in \cite{Kach} in the case of k-inflation, and in the slow roll and small speed of sound limits \footnote{To compare Eq. (\ref{fnllim}) with Eq. (4.39) of \cite{Kach} one needs: $\lambda_\text{ours}/c_s^2 = 6 \lambda_\text{theirs}/\Sigma$, where $\Sigma =  X \partial_X P + 2 X^2 \partial_X^2 P$ and $\lambda_\text{theirs} = X^2 \partial_X^2 P + \frac{2}{3} X^3 \partial_X^3 P$.}.  If one is interested in a finite speed of sound, one then simply replaces $1/c_s^2 \rightarrow (1/c_s^2 - 1)$ in the second term of Eq. (\ref{fnllim}).

  So far we have assumed that the speed of sound is small. However, another way of getting large non-gaussianities is to make $\lambda$ large. In this case,  $f_{NL}$ will be dominated by the first term in Eq. (\ref{fnlphi}):
  \be \label{lambdalarge}\lim_{|\lambda| \rightarrow \infty} f_{NL}=   \frac{5 (2 - b) \lambda  k_1^2 k_2^2 k_3^2}{6 a^4 \tilde c_s^2 K^3 \sum_i k_i^3} \;.\ee  Note that the dependence on the speed of sound is exact in this case (see \cite{Kach} and footnote [28]).
  Therefore, in this general class of models, we can parameterise the leading contribution to $f_{NL}$ by three parameters: $c_s$, $ \lambda$ and $g$.  In the next sections we will see how observations put constraints on these parameters.

   \subsection{The Shape of Non-Gaussianities}
   Studying the shape of $f_{NL}$ amounts to calculate how it depends on the size and shape of the a triangle with sides $k_i$. To get an easier visualization, we will follow \cite{shape}, and plot $f_{NL}$ as a function of the two ratios: $x_2 = k_2/k_1$ and $x_3 = k_3/k_1$. Note that the dependence of $k_1$ drops out. 
   We will begin by studying the effects of the Gauss-Bonnet coupling. Therefore we set the extra parameter $\lambda = 0$  for the moment.

   The shape of $f_{NL}$ for $\lambda = g = 0$ is shown in figure 1.  We normalize $f_{NL}$ so that it is equal to one in the equilateral limit $k_1 = k_2 = k_3$. However, we respect the sign that follows from the definition given in the previous section. Moreover, in order to avoid overcounting configurations, we set $f_{NL} = 0$ if $x_2$ and $x_3$ do not obey the inequalities, $ 1 - x_2 \leq x_3 \leq x_2$.
    \myfig{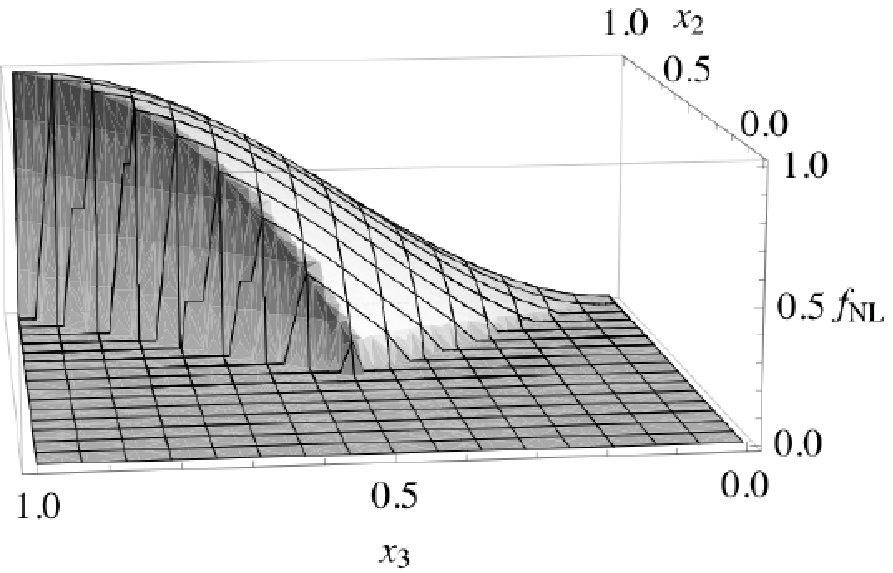}{8}{The non-gaussianity parameter $f_{NL}$ as a function of  $x_2 = k_2/k_1$ and $x_3 = k_3/k_1$ for $g = \lambda = 0$.  We have normalized $f_{NL}$ so that it is equal to one in the equilateral limit $k_1 = k_2 = k_3$. } 
      The shape shown in figure 1 is what we expect in usual k-inflation (with $\lambda = 0$). 
      
If we turn on the Gauss-Bonnet coupling, we get a slight deformation of this shape. However, it is easy to show that $f_{NL}$ is bounded even if we take $g\rightarrow \infty$. The deformation  of $f_{NL}$ is clearest if we plot the change of the normalized non-gaussianity,
\bes \Delta f_{NL} \equiv f_{NL}^\text{norm}(g)-f_{NL}^\text{norm}(g = 0)\;,\ees
where $f_{NL}^\text{norm}$ means that we have normalized $f_{NL}$ as described in the previous paragraph. In figure 2 we plot $\Delta f_{NL}$ for $g = 3$. In the next section we will see that this value of $g$ is within the current observational limits. 
  \myfig{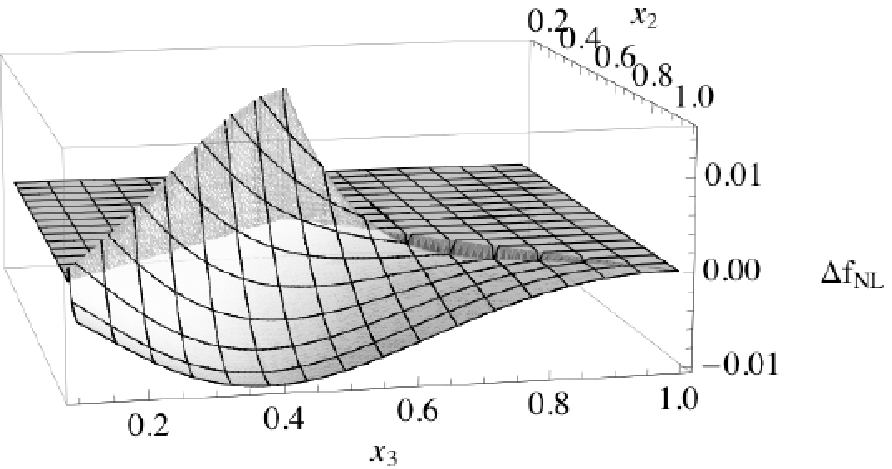}{10}{Deformation of $f_{NL}$ due to the Gauss-Bonnet coupling for $g = 3$  ($\lambda = 0$).}
  \myfig{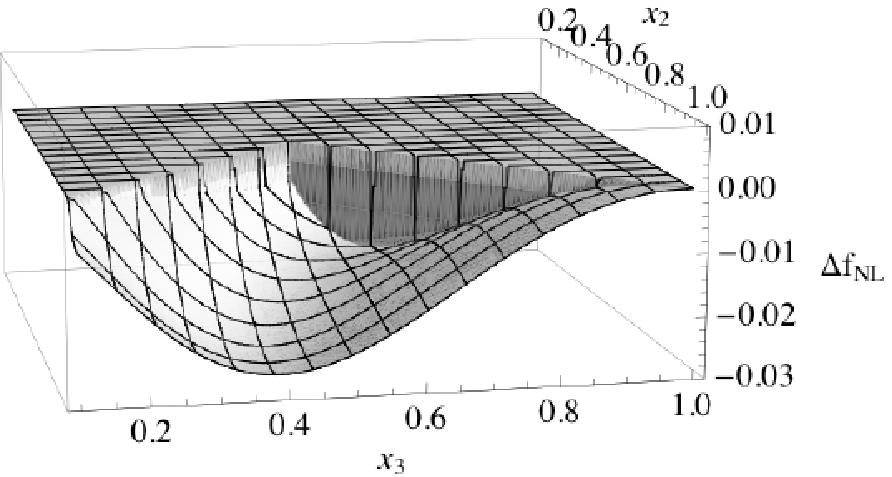}{10}{Deformation of $f_{NL}$ due to the Gauss-Bonnet coupling for $g \rightarrow \infty$ ($\lambda = 0$).}
  It is interesting that, for all values of $g$ the deformation of $f_{NL}$ obeys the bound $|\Delta f_{NL}| \lesssim 10^{-2} |f_{NL}(g = 0)| \sim 10^{-2} /c_s^2$. However, this is about the same order of magnitude as the contributions coming from the slow roll parameters which are generically of order $\sim \epsilon/c_s^2 \sim |n_s - 1|/c_s^2 \sim 10^{-2}/c_s^2$, where $n_s$ is the scalar tilt. In the most optimistic scenario, where $1/c_s^2 \sim 10^2$, these deformations will represent a change of order $|\Delta f_{NL}| \sim {\cal O}(1)$. This is about the lower end of the detectability threshold of non-Gaussianities \cite{Komatsuetc}. Therefore, we conclude that the Gauss-Bonnet coupling does not produce a measurable deformation of the shape of $f_{NL}$ from that of k-inflation. 
  
  Let us now turn our attention to the effect of the parameter $\lambda$.  To estimate the biggest possible contribution from this term, we assume $\lambda \gg 1$. Then, we plot the difference
  \bes \Delta f_{NL} \equiv  f_{NL}^\text{norm} (g = 0,\lambda\rightarrow \infty)-f_{NL}^\text{norm}(g = 0, \lambda = 0)\;,\ees
where
  \bes f_{NL}^\text{norm} (g = 0,\lambda\rightarrow \infty)= \frac{81  k_1^2 k_2^2 k_3^2}{ K^3 \sum_i k_i^3}\;.\ees
  The resulting plot is shown in figure 4. 
  \myfig{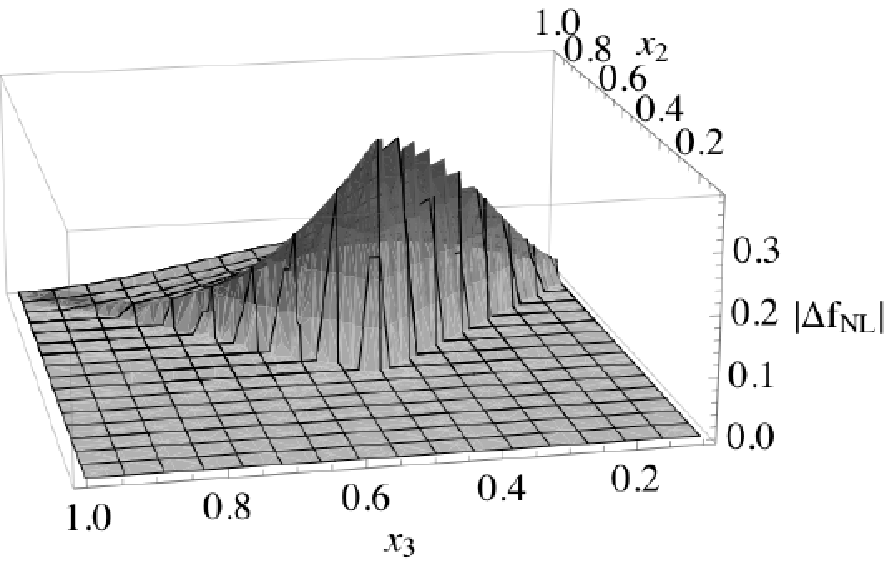}{10}{Deformation of $f_{NL}$ for  $\lambda \gg 1$. We have set $g = 0$ in this plot.}
  We can see that the deformation due to $\lambda$ is an order of magnitude larger than the rest of the deformations. Therefore, it is potentially observable. Although the plot of figure 3 is for $\lambda\rightarrow \infty$, we have found that the deformation is of order $|\Delta f_{NL}^\text{norm}| \sim {\cal O}(10^{-1})$ for $|\lambda|\gtrsim |1 - c_s^2|10^3$.

   If a large $f_{NL}$ is observed in the CMB, one could test whether the dominant effect is a small speed of sound or large $\lambda$, by looking at the deformation around the basic shape with $g = \lambda = 0$. However, such test will not put any constraints on $g$ since we have seen that its effect on the shape of $f_{NL}$ is negligible.
  Nevertheless, there is another effect of the Gauss-Bonnet coupling that we have not studied so far. In the next section we show that the Gauss-Bonnet coupling amplifies the spectrum of gravitational waves. We will see that this enhancement allows to take a smaller speed of sound, and so in an indirect way, the Gauss-Bonnet coupling can also lead to an amplification of the non-gaussianities. We will show that current WMAP data can already put constraints on the value of $g$.
  
  \section{ The Gravitational Wave Spectrum}
  
  In this section we will study the gravitational wave spectrum of the Gauss-Bonnet coupling in the slow roll limit. We will only consider the quadratic fluctuations. Similar studies have been done in \cite{hwang2} for a field with canonical kinetic term. Note that the axion coupling in the action (\ref{S}) does affect the gravitational wave spectrum. However, the effects of this term were understood in \cite{Weinberg} and \cite{hwang1}. The axion coupling introduces a helicity dependence  in the tensor power spectrum. We will ignore this term in the following.
  
  The Gauss-Bonnet term, however, has a very different effect. It allows for both a large amplitude of gravitational waves {\it and}  large non-gaussianity. This can be traced back to the fact that  $g$ can be much larger than the slow roll parameters. Moreover, as we will show below, a large $g$ with a fixed tensor/scalar ratio,  also leads to a smaller value of the speed of sound.

  The quadratic actions for gravitational waves are given by,
  \bea \label{Sg1} S^{\gamma}_\text{EH} &=& \frac{1}{8} \int e^{3 \sigma} \left[ \dot \gamma_{i j}^2 - 2(3H^2 + 2 \dot H) \gamma_{ij}^2 - e^{-2\sigma} (\nabla \gamma_{ij})^2\right]\;, \nonumber \\
S^{\gamma}_\phi &=& -\frac{1}{4} \int e^{3\sigma} P(X,\phi) \gamma_{ij}^2\;,\nonumber \\
\label{Sg3}S^{\gamma}_\text{GB} &=& \frac{1}{8} \int e^{3 \sigma} \left[ g \dot \gamma_{i j}^2  - 2 H^2(\eta_1 + g(2 - 2 \epsilon + \xi)) \gamma_{ij}^2 - e^{-2 \sigma} (\eta_1 + \xi g) (\nabla \gamma_{ij})^2\right]\;,\nonumber  \eea

In the slow roll limit, the total quadratic action simplifies to
give
\bes S^{\gamma} = \frac{1}{8} \int e^{3 \sigma}\left[ (1 + g) \dot \gamma_{ij} - e^{-2\sigma} (\nabla \gamma_{ij})^2\right]\;,\ees
where we have used the equations of motion (\ref{eom1})  and (\ref{eom2}).

The power spectrum is then easily calculated:
\bes \bra \gamma_{\vec{k}}^s \gamma_{\vec{k}'}^{s'} \ket = \frac{2 (1 + g)^{1/2} H^2}{k^3}  \delta_{s  s'}(2\pi)^3 \delta^{(3)}(\vec{k} + \vec{k}')  \equiv \frac{{\cal P}_\gamma(k)}{k^3}  \delta_{s s'} (2\pi)^3 \delta^{(3)}(\vec{k} + \vec{k}')\;.\ees
where $s$, $s'$ labels the two helicities of the graviton. Using the scalar power spectrum Eq. (\ref{Pscalar}), the tensor/scalar ratio takes the form,
\be\label{r} \boxed{r \equiv \frac{{\cal P}_\gamma}{{\cal P}_\zeta} = \frac{16 \tilde c_s g (1 + g)^{5/2}}{(2 + 3 g)^2}\;.}\ee

In the limit where the Gauss-Bonnet coupling vanishes, the dominant contribution to $r$ will come from the slow roll parameter $\epsilon$ \cite{Kach}
\be \label{rlim} \lim_{g \rightarrow 0} r = 16 c_s \epsilon\;.\ee
Therefore, we see that a small speed of sound suppresses the gravitational wave amplitude in this case. However, in the case of a non-vanishing Gauss-Bonnet coupling, we see from (\ref{r}) that $g$ can be relatively large to compensate for a small speed of sound. In this way one can produce a large amplitude of gravitational waves {\it and} a large value of $f_{NL}$.
Of course, as we discussed in the previous section, one can also produce large non-gaussianities with a large value of $\lambda$. In this case, the speed of sound does not need to be large.

 We now want to know if we can constraint the parameters $c_s$, $g$ and $\lambda$ using current CMB data.  The current limit on the tensor/scalar ratio is $r \lesssim 0.20$ \cite{Komatsu}. Therefore, from Eq. (\ref{r}) we see that $g$ cannot be too large without decreasing the value of the speed of sound. The allowed parameter space for $\tilde c_s$ and $g$ is shown in figure 5.
 \myfig{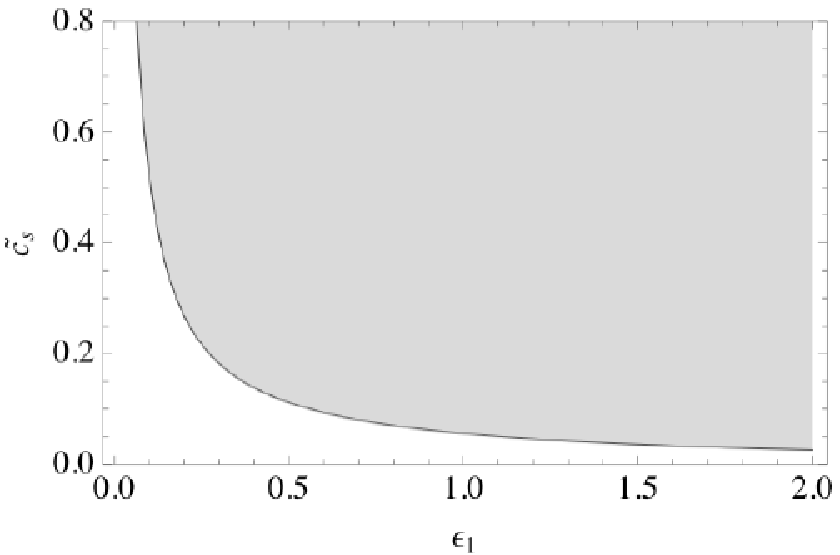}{9}{Observational constraints on $g$ and $\tilde{c}_s$. The shaded region has been ruled out by the observations. }
 
 Note that for non-trivial values of the Gauss-Bonnet coupling ($g \gtrsim 0.1$) we need a small speed of sound. This will, in turn translate to large non-gaussianities.  Moreover, a small value of $g$ will be very hard to disentangle from the contribution of the slow roll parameters. Therefore, we see that the most interesting region is that of a small speed of sound, and hence large $f_{NL}$.  This is consistent with the approximations made in the calculation of $f_{NL}$.

 We now want  to constraint $\lambda$ and $g$ combining limits on $r$ and on $f_{NL}$. 
  To do this, we eliminate $\tilde c_s$ in terms of $r$ and $g$ using Eq. (\ref{r}). We can then insert this value into $f_{NL}$,  Eq. (\ref{fnlfinal}), and write the equilateral $f_{NL}$ as a function of $r$, $g$ and $\lambda$.
In the equilateral limit, the formula for $f_{NL}$ simplifies to,
\be \label{fnleq} f_{NL}^\text{eq} = \frac{1}{\tilde c_2^2}\left[ \frac{5  \lambda (1 + 2 g)^2(2 + 3 g)}{486 (1 + g)^3} - \frac{5 \left(2 \epsilon _1+1\right) \left(\epsilon _1 \left(\epsilon _1 \left(\epsilon _1 \left(72 \epsilon
   _1+379\right)+690\right)+516\right)+136\right)}{1296  \left(\epsilon _1+1\right){}^4 \left(3 \epsilon _1+2\right)} \right]\;.\ee
where,
\bes \tilde c_s = \frac{r (2 + 3 g)^2}{16  g (1 + g)^{5/2}}\;.\ees
Current WMAP bounds on the equilateral non-gaussianities are roughly $|f_{NL}^\text{eq}| \lesssim 250$ \cite{Komatsu}.  Given these constraints, we show the allowed parameter space for $g$ and $\lambda$  in figure 6.
\myfig{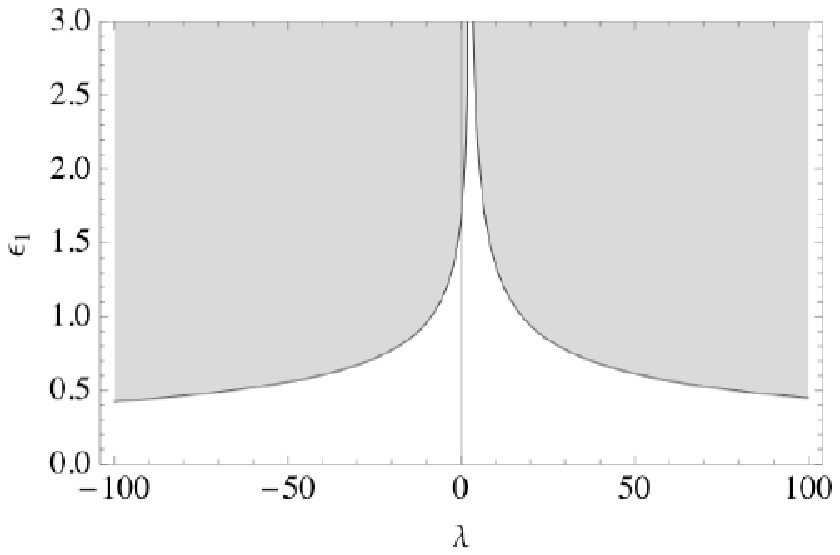}{9}{Observational constraints on $g$ and $\lambda$. The shaded region has been ruled out by the observations. }

One can see a degeneracy point around $\lambda \approx 2$ where one can take large values of $g$. This is due to the fact that we can have cancelations between the two terms in brackets in Eq. (\ref{fnleq}). However, this is a very fine tuned situation. 
Nevetheless, we see that most of the parameter space for $g$ is quite well constrained by current observations. 
This is specially true for models with large $|\lambda|$.

The bound on $\lambda$ is, however, not very good. The absolute limit comes from taking $g \rightarrow0$ and using the limit of $r$ given in Eq. (\ref{rlim}). The equilateral $f_{NL}$ simplifies to
\be \lim_{g \rightarrow 0} f_{NL}^\text{eq} = \frac{320 \epsilon ^2 (4 \lambda -51)}{243 r^2}\;.\ee
Since $\epsilon$ is enters linearly in the scalar tilt, one has that $\epsilon \lesssim 10^{-2}$ given current WMAP observations. Using the limit on equilateral $f_{NL}$ and $r$ given above, one gets a rough bound
\be |\lambda| \lesssim 10^4\;.\ee

So far we have concentrated in giving bounds on $g$ and $\lambda$. However, one can ask if it is possible to break the degeneracy between these two parameters and put more precise constraints on a non-zero Gauss-Bonnet coupling. The most optimistic scenario would be to detect a large $f_{NL}$ and the deformation  due to $\lambda$ described in the previous section.  In this case one would confirm that $|\lambda| \gtrsim |1 - c_s^2| 10^3$. This, combined with limits on $r$ (see figs. 5 and 6) will put tight constraints on $g$. 
Nevertheless, in most models  $\lambda$ is not an independent parameter. In this case we can put more precise constraints on the Gauss-Bonnet coupling. We illustrate this point in the next section for the case of DBI inflation.

\section{An Example: DBI Inflation}
So far we have discussing the constraints on the Gauss-Bonnet coupling in a model independent way. However,  in order to break the degeneracy between the parameters $g$ and $\lambda$ we need to consider particular models. In this section we will study DBI inflation \cite{DBI}. This is a string-inspired model where the inflaton encodes the position of a D-brane on a warped compactification. 
In this model, the function $P(X,\phi)$ takes the form
\bes P(X,\phi) = -f(\phi)^{-1} \sqrt{1 - 2 X f(\phi)} + f(\phi)^{-1} - V(\phi)\;,\ees
where $f(\phi)$ is related to the warp factor of the compactification.

Then, it is easy to show from the definition of $\lambda$, Eq. (\ref{lambda}), that 
\bes \lambda_\text{DBI} = 3\;.\ees
In this model we do not have any degeneracy between the parameters and we can constraint $g$ directly. In figure 7 we plot the the absolute value of the equilateral $f_{NL}$ as a function of $g$ and the tensor/scalar ratio $r$. We see that in DBI inflation,  the Gauss-Bonnet coupling is constrained as 
\be\label{bound} \boxed{ g_\text{DBI} \lesssim 3\;.}\ee
Moreover, we see that there is a very interesting region of parameter space where we can have both large $f_{NL}$ and a large amplitude of gravitational waves. 
\myfig{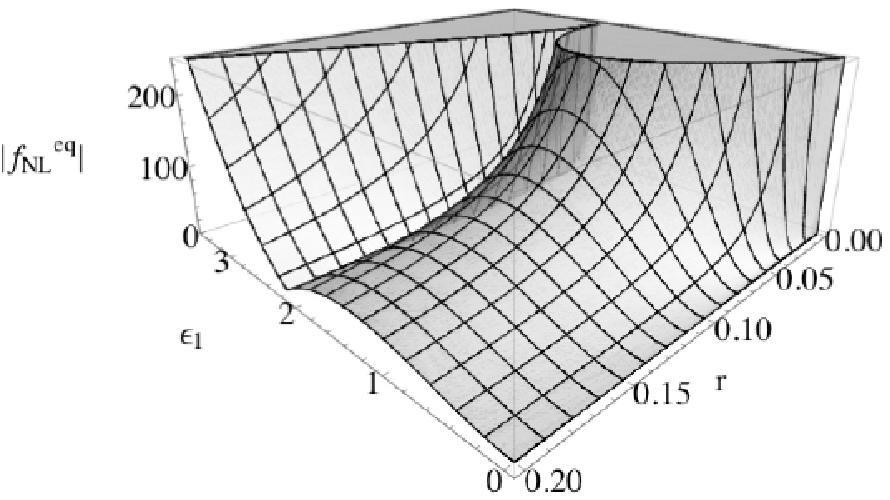}{10}{Equilateral $f_{NL}$  for DBI inflation, as a function of the tensor/scalar ratio $r$ and the Gauss-Bonnet coupling strength $g$. The observationally allowed region is for  $r \lesssim 0.20$ and $|f_{NL}^\text{eq}| \lesssim 250$. }

So far we have assumed that the slow roll parameters are very small compared to any other scale in the problem.  This will translate in  some constraints on the potentials $V$, $V_1$ and $f$. In the slow roll limit, the equations of motion (\ref{eom1}) and (\ref{eom2}) reduce to,
\bea \label{dbi1} \frac{\dot\phi^2}{H^2} &=& c_s g\;,\\
 \label{dbi2} \frac{1}{f(\phi)}\left(\frac{1}{c_s} - 1\right) + V(\phi) &=& 3 H^2 (1 + g)\;,\eea
 where,
 \be c_s  = \sqrt{1 - 2 X f}\;.\ee
 To solve these equations, one first solves for $\dot\phi$ from Eq. (\ref{dbi1}) and substitute it on Eq. (\ref{dbi2}). This gives a Hubble parameter that is a function of the scalar field $H = H(\phi)$. We then substitute this back in (\ref{dbi1}) to solve for $\phi$ as a function of time. We will not do this in details, since we are only  interested in deriving general conditions on the potentials. 
 
  In the limit of a small speed of sound, one has $\dot\phi^2 \approx 1/(2 f)$.  One can then show that
  \bea\label{Hdbi} H^2(\phi) &\approx& \frac{V(\phi)}{3 + 2 g(\phi)}\;,\\
  \label{epsilon1dbi}g(\phi) &\approx& \frac{8 V_1'(\phi) H(\phi)}{\sqrt{f(\phi)}}\;,\\
  \label{csdbi}c_s(\phi) &\approx& \frac{1}{8 V_1'(\phi) H^3(\phi) \sqrt{f(\phi)}}\eea
  From Eqs. (\ref{Hdbi}) and (\ref{epsilon1dbi}), we see that $H$ obeys a cubic equation. However, we do not need to solve this equation explicitly to derive the slow roll conditions. Using Eqs. (\ref{Hdbi}) - (\ref{csdbi}) along with (\ref{slowroll1}) and (\ref{slowroll2}), one can derive the following relations between the slow roll parameters:
  \bea \epsilon &\approx&  - \frac{ V'(\phi) }{2 V(\phi)} \frac{\dot\phi}{H}+ \frac{g}{3+ 2 g} \frac{\dot g}{H g}\;,\\
  \frac{\dot g}{H g} &=&  \frac{V_1''(\phi) }{V_1'(\phi)} \frac{\dot\phi}{H} - \epsilon + \xi \;,\\
  \xi &\approx&-\frac{ f'(\phi) }{f(\phi)} \frac{\dot\phi}{H}\;,\\
  \frac{\dot c_s}{H c_s} &\approx& - \frac{\dot\phi}{H} \left( \frac{V_1''(\phi)}{V_1'(\phi)} + \frac{ f'(\phi)}{2 f(\phi)}\right) + 3 \epsilon\;.
  \eea

   Our results have assumed that all slow roll parameters are much smaller than $\dot\phi/H$. Therefore, to be consistent we should impose the conditions:
   \be \label{cond}  \frac{f'(\phi)}{f(\phi)} \ll1\;,\;\;\; \frac{V_1''(\phi)}{V_1'(\phi)}\ll 1\;,\;\;\; \frac{V'(\phi)}{V(\phi)}\ll 1\;,\;\;\; \frac{V''(\phi)}{V(\phi)}\ll 1\;,\ee
where the last condition follows from $\dot \epsilon/ (H\epsilon) \ll 1$.  

For an AdS throat, the warp factor is $f(\phi) = \lambda/\phi^4$, where $\lambda$ is the `t-Hooft coupling of the dual gauge theory, and $\lambda \gg 1$ \cite{DBI}. Therefore, the first condition in (\ref{cond}) gives, $\phi \gg 1$.  In other words, the D3-brane must be in the UV region of the AdS throat. Note that this is still consistent with the bounds studied in \cite{DBI}, so that one can ignore the backreaction on the geometry. 

If we want to be in the interesting window of large $f_{NL}$ and observable gravitational waves, we need $g  \sim {\cal O}(1)$. In general, this condition will require a very steep potential $V_1$. In fact, using Eq. (\ref{r}) for the tensor/scalar ratio, along with Eq. (\ref{dbi1}) for $\dot\phi$, one can show that in this regime $V_1'(\phi) \sim 10^{10}$.  Since, for the AdS throat, one is interested in $\phi \gg1$, one could perhaps realize such large value of $V_1'(\phi)$ with a power-law potential $V_1(\phi)\sim\phi^n$. Note that the slow roll conditions in (\ref{cond}) will automatically be satisfied for large values of $\phi$. Wether such scenario can be realized in a controlled string theory construction, is beyond the scope of this paper. 

\subsection{Comments on the Lyth Bound}
We have seen that a Gauss-Bonnet coupling enhances the amplitude of gravitational waves. It was shown in  \cite{Lyth} that, in the context of slow roll inflation, an observable amplitude of gravitational waves would require a ultra-planckian displacement of the inflaton. This is known as the Lyth bound.  Quite generally, we can write the bound as
\be\label{lyth} \Delta\phi > | \dot\phi/H | \Delta {\cal N}\;,\ee
where $\Delta {\cal N}$ is the number of e-foldings in which the scales of interest in the CMB today exit their horizon. This is usually taken as $\Delta {\cal N}\approx 4.6$ \cite{Lyth}. Using Eqs. (\ref{dbi1}) and (\ref{r}) in (\ref{lyth}), one can write the Lyth bound in terms of the tensor/scalar ratio and the Gauss-Bonnet coupling:
\bea \label{lyth1} \Delta\phi \gtrsim 0.36183 \frac{3 g+2}{\left(g+1\right){}^{3/4} (g
   \left(6g+7\right)+2)^{1/4}} \sqrt{\frac{r}{0.07}}  \rightarrow \frac{0.693566}{g^{1/4}} \sqrt{\frac{r}{0.07}}\;,\eea
   where in the last step we have shown the limit of large $g$. Moreover, $r \approx 0.07$ is considered to be the lowest limit of detectability for gravitational waves \cite{Lyth}. 
   
  We can see from Eq. (\ref{lyth1}) that one can, in principle, violate the Lyth bound if we have a large Gauss-Bonnet coupling.  However, a large value of $g$, with fixed $r$ requires a very small speed of sound (see Eq. (\ref{r})). We have seen that current WMAP limits on $f_{NL}$ already constraint the value of the Gauss-Bonnet coupling to Eq. (\ref{bound}). With this value we get $\Delta \phi \gtrsim 0.5$, which is roughly the same bound as in slow roll inflation. Nevertheless, if we are working with an AdS throat, we need $\phi \gg1$ and so $\Delta\phi/\phi \ll1$. So we see that the fractional change in the scalar field is very small. This means that, in order to make CMB predictions within this class of models, we only need to know the potentials in the large $\phi$ limit.   
   It would be interesting to see wether one can violate the bound in some other model with a Gauss-Bonnet coupling which allows $\phi \sim 1$.

\section{Conclusion}
In this article we have studied WMAP constraints on modifications of gravity due to a Gauss-Bonnet coupling in single-field inflation. This is the most general modification of gravity that leads to second order equations of motion, and that also affects the spectrum of scalar fluctuations.
We showed that in the slow roll limit, and for a very general class of models with action (\ref{S}), a large $f_{NL}$ can be written in terms of three parameters: $c_s$, $\lambda$ and $g$ the Gauss-Bonnet coupling.

We found that the Gauss-Bonnet term has little effect on the shape of non-gaussianities, but it amplifies the tensor power spectrum. Thus, given current limits on the tensor/scalar ratio $r$, we found that large values of the Gauss-Bonnet coupling would require small values of the speed of sound, and hence large non-gaussianities. Using current WMAP limits on $r$ and $f_{NL}$ we were able to constraint the parameter space of such models. 

To give better constraints on $g$ we studied a particular model: DBI inflation. In this case we obtained a precise bound on this coupling, Eq. (\ref{bound}). Moreover, we saw that in this model, a non-zero Gauss-Bonnet couplings leads to an interesting observational window with both large non-gaussianities and a large amplitude of gravitational waves. We also studied how the conditions for the smallness of the slow roll parameters translate to constraints on the scalar potentials. Possible violations of the Lyth bound were also studied. We found that the bound can be violated for large values of $g$. However, for DBI inflation one has an observational restriction of $g \lesssim 3$, and so the bound is roughly $\Delta \phi \gtrsim 0.5$. Nevertheless, we found that for an AdS throat the fractional change in the scalar field is very small: $\Delta\phi/\phi \ll 1$.
It would be interesting to study the higher curvature corrections to the DBI action, to see if one can realize a Gauss-Bonnet driven inflation in a controlled way. Some of these  corrections were derived in \cite{Bachas}.

Another interesting aspect of the Gauss-Bonnet term is that scalar fluctuations can exist in a de-Sitter background. Moreover, they have a relativistic dispersion relation, unlike Ghost Inflation \cite{ghost}. This matches with the new de-Sitter limit found in \cite{Senatore} using the effective action of the scalar fluctuations.  One might wonder if there are other de-Sitter limits of inflation. In \cite{Senatore}, it was argued that the answer is negative as any such limits will not make sense as an effective field theory for the fluctuations. However, the authors of \cite{Senatore} only considered models with one scalar degree of freedom. If we ask about other modifications of gravity, it is well known that  one needs to add more degrees of freedom to the theory. 
It would be interesting to put constraints in other types of modified gravity using large non-gaussianities.

\section*{Acknowledgments}
I am very grateful to  Niayesh Afshordi for all his insight during this project. I also thank Mark Wyman for reviewing the manuscript. I would also like to acknowledge many interesting conversations with Justin Koury, Claudia de Rham,  	Ghazal Geshniziani, Andrew Tolley, Aninda Sinha,  Robert McNees   and Xiao Liu. Research at
Perimeter Institute is supported by the Government of Canada through
Industry Canada and by the Province of Ontario through the Ministry
of Research \& Innovation.

\end{document}